\documentclass[showkeys]{revtex4}
\usepackage{graphicx}
\usepackage{amsmath}
\usepackage{algorithm,algorithmic}

\newcommand{\TO}{\textbf{to}~}

\begin{document}
\title{
Huge-scale Molecular Dynamics Simulation of Multibubble Nuclei
}

\author{Hiroshi Watanabe\footnote{hwatanabe@issp.u-tokyo.ac.jp}$^1$, Masaru Suzuki$^2$, and Nobuyasu Ito$^3$}

\affiliation{
$^1$The Institute for Solid State Physics, The University of Tokyo,
Kashiwanoha 5-1-5, Kashiwa, Chiba 277-8581, Japan
}

\affiliation{
$^2$ Department of Applied Quantum Physics and Nuclear Engineering, Kyushu
University, 744 Motooka, Nishi-ku, Fukuoka 819-0395, Japan
}

\affiliation{
$^3$ Department of Applied Physics, School of Engineering,
The University of Tokyo, Hongo, Bunkyo-ku, Tokyo 113-8656, Japan
}

\begin{abstract}
We have developed molecular dynamics codes for a short-range interaction potential
that adopt both the flat-MPI and MPI/OpenMP hybrid parallelizations on the basis of
a full domain decomposition strategy.
Benchmark simulations involving up to 38.4 billion Lennard-Jones particles were performed on
PRIMEHPC FX10, consisting of 4800 SPARC64 IXfx 1.848 GHz processors, at the Information Technology Center of
the University of Tokyo, and a performance of 193 teraflops was achieved, which corresponds to a 17.0\% execution efficiency.
Cavitation processes were also simulated on PRIMEHPC FX10 and SGI Altix ICE 8400EX
at the Institute of Solid State Physics of the University of Tokyo, which involved 1.45 billion and 22.9 million particles, respectively.
Ostwald-like ripening was observed after the multibubble nuclei.
Our results demonstrate that direct simulations of multiscale phenomena involving phase transitions
from the atomic scale are possible and that the molecular dynamics method is 
a promising method that can be applied to petascale computers.
\end{abstract}


\maketitle

\section{Introduction}

Most recent computers consist of multiple nodes, with each node usually having multiple processors.
Additionally, not only state-of-the-art computer but commodity cluster systems adopt hierarchical
memory models, \textit{i.e.}, shared memory within a node and distributed memory across the nodes.
Corresponding to the different memory models, there are two programming models,
the Message Passing Interface (MPI) library and OpenMP, which are the de facto standards for threading
distributed and shared memory models, respectively.
To achieve better performance on systems with such a heterogeneous memory model,
programmers have to choose their strategies for parallelization from the following three possible approaches;
fully distributed memory models, global address space models, and a combination of distributed and shared memory models.
The fully distributed memory models are achieved by the so-called flat-MPI strategy, \textit{i.e.},
shared memory in a node is treated as distributed memory and only the MPI library is used for both intra- and internode communication.
This strategy reduces development costs, since a programmer can ignore the heterogeneity of the memory model in the system.
Also, the flat-MPI approach usually achieves acceptable performance since the MPI library is usually
optimized so that it can utilize a shared memory model within a node.
The global address space models allow programmers to use a virtually global memory address space,
which is distributed over nodes. Using partitioned global address space (PGAS) languages,
programmers can control the affinity of the memory space to threads.
There are many languages that adopt this programming model, for example,
Unified Parallel C~\cite{UPC}, Co-Array Fortran~\cite{Co-Array-Fortran},
Titanium~\cite{Titanium}, XcalableMP~\cite{XcalableMP}, and so forth.
While these languages reduce the development cost of software, users have to utilize their features to the full
to obtain sufficient performance for large-scale computations.
Finally, the combination of distributed and shared memory models is usually achieved by
using the MPI library for intranode communication and OpenMP for intranode communication.
Although it seems natural to adopt the distributed memory model across nodes and the shared memory model
within each node, the flat-MPI approach usually gives better performance than the hybrid approach.
However, one cannot always adopt the flat-MPI approach since it
requires more resources such as the total memory, especially for massively parallel simulations.
Many comparisons between the flat-MPI approach and the OpenMP/MPI hybrid approach have been carried out.
However, OpenMP/MPI hybrid programming is usually implemented by domain decomposition with the MPI library
and loop-level decomposition with OpenMP. Even if the total systems
of two schemes are identical, the computations applied to each core are different.
This makes difficult to compare the efficiency between the two schemes, since
it is difficult to find reasons for the difference in performance.

In this paper, we describe a pseudo-flat-MPI implementation of molecular dynamics (MD) simulations
that allows us to compare the performance between flat-MPI and a hybrid using 
the same program. We adopt the domain decomposition strategy for both intra- and internode
parallelizations. We also focus on how to perform product runs using the hybrid parallelized MD,
since the development cost of implementing codes for product runs can be compared with that
of implementing benchmark codes. 
The scaling analysis of benchmark simulations involving up to 38.4 billion Lennard-Jones (LJ) particles
is performed on PRIMEHPC FX10, consisting of 4800 SPARC64 IXfx 1.848 GHz processors, at the Information Technology Center of
the University of Tokyo. To achieve better performance on FX10,
we optimize the simulation kernel and achieve a performance of 193 teraflops, which corresponds to a 17.0\% execution efficiency.
We also perform substantial simulations of cavitation phenomena involving
up to 1.45 billion particles on two different platforms,
FX10 and SGI Altix ICE 8400 EX (Intel Xeon 2.93 GHz processors), at the Institute of Solid State Physics of the University of Tokyo.

This paper is organized as follows.
In Sec.~\ref{sec_method}, details of our method are described including the parallelization.
Benchmark results are given in Sec.~\ref{sec_benchmark}
and cavitation phenomena are studied in Sec.~\ref{sec_cavitation}.
Finally, Sec.~\ref{sec_summary} is devoted to a summary and discussions of this study.

\section{Method} \label{sec_method}

\subsection{Basic Algorithms}

We perform classical MD simulations of LJ potentials with truncation.
Here, we briefly describe the algorithms for short-range MD. Refer to a previous paper 
for detailed descriptions~\cite{mdnote}. The two-body potential of truncated LJ particles is given by~\cite{Spotswood1973}
\begin{equation}
V(r) = 
\displaystyle 4 \varepsilon \left[
\left( \frac{\sigma}{r} \right)^{12} -
\left( \frac{\sigma}{r} \right)^{6} +
c_2 \left( \frac{r}{\sigma} \right)^2+ c_0 \right], \label{eq_lj_cutoff}
\end{equation}
with particle distance $r$, well depth $\varepsilon$, and atomic diameter $\sigma$.
The two additional coefficients, $c_2$ and $c_0$, are determined so that
the potential and force become zero at the cutoff length $r_\mathrm{c}$, \textit{i.e.}, $V(r_\mathrm{c}) = V'(r_\mathrm{c}) = 0$.
In the following, we use physical quantities reduced by $\sigma$, $\varepsilon$, and the Boltzmann constant $k_\mathrm{B}$,
\textit{i.e.}, the length scale is measured using the unit of $\sigma$, and so forth.
We set the cutoff length to $r_\mathrm{c} = 2.5$ for benchmark simulations
and to $r_\mathrm{c} = 3.0$ for simulations of cavitation.
Since interactions between particles are short-range, it is necessary to find particle pairs
separated by a distance of them is less than the cutoff length.
We adopt a grid algorithm~\cite{Allen,Knuth1973,Grest1989,Beazley1994} (also called the linked-list method~\cite{Quentrec1975, Hockney1981}) to find the interacting particle pairs.
Since the computational cost of constructing a list of interacting particle pairs is high, 
we adopt the bookkeeping method to reduce the cost of constructing the list~\cite{Verlet1967},
which allows us to reuse the same pair list for several time steps by registering pairs within some
length which is longer than the cutoff length.
Several steps after the construction of the pair list, there may be particle pairs
separated by a distance less than the cutoff length that are not registered in the list.
If such pairs exist, the simulation fails since the total energy will not be conserved.
Therefore, we have to check the validity of the list in every time step.
For this, we adopt the dynamic upper time cutoff (DUTC) method~\cite{Isobe1999}.
The validity of the list is checked using the maximum velocity and the displacements of the particles~\cite{mdnote}.
This validity check involves the global synchronization between all processes.

\begin{figure}[htb]
\begin{center}
\includegraphics[width=0.8\linewidth]{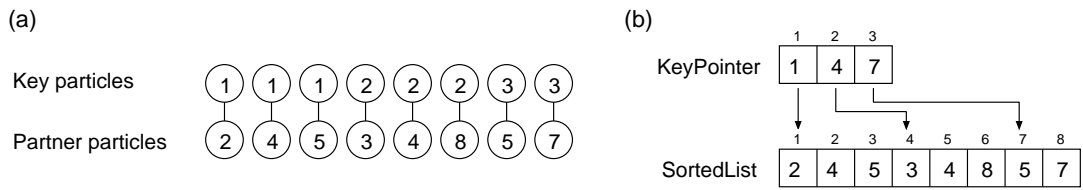}
\end{center}
\caption{
(a) Sorted pair list. The pair list is sorted by the indices of \textit{key} particles.
(b) Data format of the sorted pair list. A sorted pair list is expressed by two arrays; KeyPointer and SortedList.
The first partner of \textit{key} particle $i$ is stored in SortedList[KeyPointer[i]],
the second is stored in SortedList[KeyPointer[i]+1], and so forth.
}
\label{fig_sortedlist}
\end{figure}

After constructing a pair list, we sort the list to reduce the number of accesses to the memory (see Fig.~\ref{fig_sortedlist}).
Consider a particle pair $(i, j)$, where $ i< j$. We call the particle of index $i$ the \textit{key} particle
and the other the \textit{partner} particle. We sort the particle pairs by the indices of the \textit{key} particles.
Then the data of the  \textit{key} particle is stored in CPU registers, and the amount of 
memory access is decreased since data fetching and storing are performed only for the 
data of \textit{partner} particles.

\subsection{Parallelization}

\begin{figure}[htb]
\begin{center}
\includegraphics[width=0.95\linewidth]{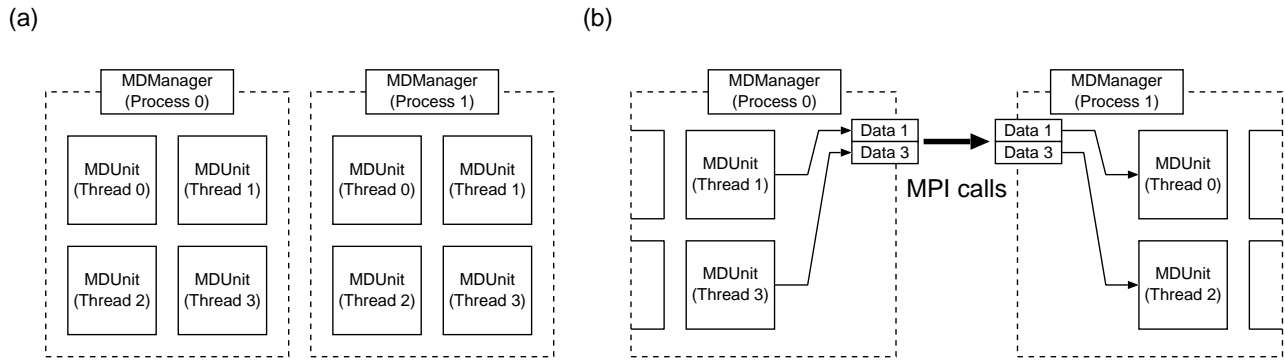}
\end{center}
\caption{
Schematic illustration of the pseudo flat-MPI approach. A two-dimensional system is shown for convenience.
There are two MPI processes,  each containing four OpenMP threads.
(a) {\ttfamily MDManager} and {\ttfamily MDUnit} represent an MPI process and an OpenMP thread, respectively.
Each domain is assigned to an {\ttfamily MDUnit}, and an {\ttfamily MDManager} takes responsibility for their communication.
(b) Communication between processes. Since a platform may not support MPI calls from a multithread environment,
data should be packed before being sent by a sender process, and the data should be distributed to destinations by a receiver process.
}
\label{fig_comm}
\end{figure}

Although there are several strategies for the parallelization of MD~\cite{Plimpton1995}, the domain decomposition method
is simple and generally scalable for a huge number of processes.
It is also the most efficient for particle systems with short-range interactions 
if the interaction length is sufficiently shorter than the size of the domain assigned to each process.
Domain decomposition parallelization is usually implemented by the flat-MPI approach.
This approach allows a programmer to consider only a single programing model, and therefore, 
the development cost is reasonable for huge scale computations.
However, limited resources, usually a lack of memory, may prevent the adoption of the flat-MPI programming model.
Thus, the hybrid programming model is required to reduce the number of MPI processes.
One method of implementing hybrid parallelized MD is to adopt domain decomposition
for internode communication with the MPI and particle decomposition for intranode communication with loop-level OpenMP.
One advantage of this method is that the implementation can be simple.
By inserting an OpenMP directive before the loop of force calculations, one can modify a flat-MPI code
to an OpenMP/MPI hybrid code. However, there are some disadvantages of this method.
Construction of the pair list typically costs 10\% of the total computation time.
If the parallelization efficiency of the force computation within a node is perfect,
then the construction of the pair list may become a new bottleneck in simulations.
Therefore, one has to implement OpenMP parallelization of the pair list construction, as well as the force calculation.
Additionally, the codes computing observables, such as energy and pressure,
can also be a bottleneck. Therefore, a programmer has to consider all routines and implement
hybrid versions of them if necessary. This pushes up the development cost.

In this study, we adopt a pseudo-flat-MPI approach, \textit{i.e.}, 
we adopt a full domain decomposition strategy 
for both intra- and internode parallelizations by using thread IDs as ranks of the MPI~\cite{Berger2005}.
The advantages of the pseudo-flat-MPI approach as follows, with details will be described later.
\begin{description}
\item[Simple implementation:] If one already has a flat MPI code, it is straightforward to modify it
to a pseudo-flat-MPI code.
\item[Flexibility:] It is easy to change the number of threads per process without rebuilding the program.
\item[Affinity with CPU-level tuning:] For the naive loop-level OpenMP paradigm, one has to consider
OpenMP parallelization and SIMD (single instruction multiple data) optimization simultaneously.
The pseudo-flat-MPI paradigm allows a programmer to consider only SIMD optimization.
\item[Memory locality:] For systems with Non-Uniform Memory Access (NUMA) architecture, a programmer has to consider
memory locality, \textit{i.e.}, which data should be assigned to which CPU core.
In the pseudo-flat-MPI paradigm, the memory locality is automatically satisfied as for the flat-MPI paradigm.
\end{description}

In the following, we describe our implementations in C++ language since we have written our codes in C++.
However, similar approaches are possible in C or Fortran 77 or other non object-oriented languages.
We define two classes; \verb|MDManager| and \verb|MDUnit|.
An \verb|MDUnit| class represents an OpenMP thread (see Fig.~\ref{fig_comm}~(a)).
The system is divided into small domains, and one domain is assigned to one instance of the \verb|MDUnit| class,
\textit{i.e.}, the instance of the \verb|MDUnit| class is the owner of the domain and is responsible for updating
the momenta and coordinates of particles in the domain.
An \verb|MDManager| class represents an MPI process.
It manages at least one instance of \verb|MDUnit| and is responsible for communication between domains.
Sample codes for the initialization of \verb|MDUnit| are as follows.
\begin{verbatim}
MDUnit *mdp;
std::vector <MDUnit *> mdv;
#pragma omp parallel shared(mdv) private(tid,mdp)
{
  tid = omp_get_thread_num();
  mdp = new MDUnit(tid);
#pragma omp critical
  mdv.push_back(mdp);
}
\end{verbatim}
Since variables are initialized in each constructor of \verb|MDUnit|, the memory locality is automatically satisfied.
The number of threads per process corresponds to the number of instances of \verb|MDUnit| managed by an instance of \verb|MDManager|.
If one  \verb|MDManager| has only one instance of \verb|MDUnit|, then the computation is performed in the manner of
the flat-MPI paradigm; otherwise it is performed in the manner of the hybrid. The number of threads per process is easily changed 
by changing the value of the \verb|OMP_NUM_THREADS| environment variable.
A sample code for calculating force is as follows.
\begin{verbatim}
#pragma omp parallel for schedule(static)
for(int i = 0; i < num_threads; i++) {
  mdv[i]->CalculateForce();
}
\end{verbatim}
Here, \verb|num_threads| is the number of threads per process.
The implementation of the \verb|MDUnit| class does not depend on whether 
it is called by a thread or a process except is the case of communication. Therefore, it is easy to 
modify an existing flat-MPI code to a pseudo-flat-MPI code.

Communication involving hybrid parallelization should be considered since
there are several levels of support for a call of MPI routines in a multithread environment.
The specifications of MPI 2.2 define the following four levels of support~\cite{MPI_spec}.
\begin{itemize}
\item \verb|MPI_THREAD_SINGLE:| Each process has only one thread.
\item \verb|MPI_THREAD_FUNNELED:| The process may be multithread, but the
application must guarantee that only the main thread makes MPI calls.
\item  \verb|MPI_THREAD_SERIALIZED:| The process may be multithread and any thread
may make MPI calls. However, the application must guarantee only one call at a time,
\textit{i.e.}, consecutive MPI calls from different threads should be serialized.
\item \verb|MPI_THREAD_MULTIPLE:| Any thread may make MPI calls without any restrictions.
\end{itemize}
If the system supports \verb|MPI_THREAD_MULTIPLE|, the implementation of 
communication in the pseudo-flat-MPI becomes identical to that in the flat-MPI.
Each instance of \verb|MDUnit| can communicate with others as if it is a process while it is a thread.
However, most platforms do not support \verb|MPI_THREAD_MULTIPLE|,
for example, the support level of both sites we use in this study is \verb|MPI_THREAD_SERIALIZED|. 
With the straightforward implementations of communications for the pseudo-flat-MPI,
two threads belonging to the same process may call \verb|MPI_Sendrecv| simultaneously
which is inhibited for \verb|MPI_THREAD_SERIALIZED|.
Therefore, a programmer has to consider intra- and internode data exchanges explicitly.
There are several strategies to treat communication between threads.
 One is a data packing strategy which works for \verb|MPI_THREAD_FUNNELED| and higher supporting levels.
The data required from other processes should be packed before being sent by a sender process
and they are distributed by a receiver process (see Fig.~\ref{fig_comm}~(b)).
This two-level communication is also required for routines to calculate observables.
Therefore, one has to rewrite all codes to calculate observables even if one has already developed 
parallelized ones for a flat-MPI program.
This is one of the disadvantages of the pseudo-flat-MPI approach.
Note that one does not have to pack the data for communication if the system
supports \verb|MPI_THREAD_SERIALIZED|. 
One can use memory copy for threads in a same process and using MPI calls for ones
belonging to different processes, but it is simpler to adopt the data packing strategy
for both  \verb|MPI_THREAD_FUNNELED| and \verb|MPI_THREAD_SERIALIZED|. 
Use of the non-blocking communications may be one of candidates.
However, we avoid using non-blocking communications for two reasons.
First, our simulation has relatively coarse granularity, and therefore, 
overlapping of communication and computation is unnecessary.
The other reason is that the non-blocking communications are
resource consuming compared with the blocking communications.
Amount of consumption of resources, such as memory, is
important especially for huge-scale computations.
Therefore, we adopt the data packing strategy even for \verb|MPI_THREAD_SERIALIZED|.

In the present manuscript, we do not consider heterogenous computation
such as systems with heterogenous architectures or general-purpose
computing on graphics processing units (GPGPUs).
There are several attempts to perform MD simulations of Lennard-Jones particles
on heterogenous environments.
Germann et al.~port the  molecular dynamics code 'SPaSM'~\cite{Beazley1994} to the hybrid supercomputer
Roadrunner which contains different CPU architecture, AMD Opteron and IBM PowerXCell~\cite{Germann2009}.
They presented two strategies, `evolutionary' and `revolutionary' approaches.
The `evolutionary' port is the strategy to obtain the benefit from the computational power of
IBM PowerXCell with minimal modification. They offloaded the force calculations
from Opteron to Cell, and obtained twofold speedup.
The `revolutionary' port is a kind of accelerator-hosted model, \textit{i.e.},
most of computations are taken up by Cell. Thanks to lots of Cell-specific optimizations,
they achieved speedup more than tenfold compared with Opteron-only calculation.
Another example is porting LAMMPS~\cite{Plimpton1995} to utilize the system with GPGPUs~\cite{Brown2011}.
They put some part of computational task from CPUs to accelerators.
Since it is difficult to know the ideal fraction of tasks between CPU and accelerators,
they implemented dynamic load balancer to achieve it.
They also reported that not only force calculation but also building pair lists
should be performed on accelerators to achieve speed up, since the building pair lists
on CPU became the bottle neck of simulations if accelerators perform only force calculations.
The pseudo-flat-MPI approach described in the present manuscript 
may work on such heterogenous environment, but considering the previous works,
it will require substantial effort to utilize the computational power of such systems to the full.

We do not consider any load balancing, either.
The dimensions of each domain assigned to a thread are identical among domains
and do not change throughout the simulations.
Although load imbalance may decrease the efficiency of simulations, the decrease is not serious for our simulations.
The reason for this is described in Sec.~\ref{sec_loadimbalance}.

\section{Benchmark Results} \label{sec_benchmark}

\subsection{Details of Platforms}

\begin{table}[tb]
\begin{tabular}{cccccc}
\hline
Facility &  System                              &  CPU                                           &  Cores  &  Memory & Nodes  \\\hline
ITC         & PRIMEHPC FX10   &   SPARC64 IXfx (1.848  GHz)   & 16                   & 32 GB     & 4800          \\
ISSP       & SGI Altix ICE 8400EX    &  Intel Xeon X5570 (2.93 GHz) & 8             &   24 GB  & 1920            \\\hline
\end{tabular}
\caption{ Summary of
(ITC) PRIMEHPC FX10 at the  Information Technology Center of the University of Tokyo and
(ISSP) SGI Altix ICE 8400EX at the Institute for Solid State Physics of the University of Tokyo.
The facility, name, CPU, number of cores per node, memory per node,
and number of nodes in the system are shown for each system.
Although the total number of nodes at ISSP is 1920, we only use 128 nodes in this study.
}
\label{tbl_computers}
\end{table}

We perform simulations on two platforms.
One is PRIMEHPC FX10 at the Information Technology Center (ITC), the University of Tokyo.
Each node contains a SPARC 64 IXfx 1.848 GHz processor containing 16 CPU cores.
The memory access is uniform from each core, \textit{i.e.}, this system
adopts Uniform Memory Access (UMA).
Each node contains 32 GB and the total amount of memory is 150 TB.
We used Fujitsu C/C++ Complier Ver.~1.2.1 with the compile options \verb|-Kfast -Ksimd=2 -Knoparallel -Kopenmp|.
Note that the compile options supported by the Fujitsu C/C++ complier can be different for different sites.

The other is SGI Altix ICE 8400EX at the Institute for Solid State Physics (ISSP), of the University of Tokyo.
Each node contains two Intel Xeon 2.93 GHz processors containing 4 CPU cores,
and each node adopts cache-coherent NUMA (ccNUMA).
The total system contains 1920 nodes and provides 180 teraflops.
Each node contains 24 GB and the total amount of memory is 46 TB.
We used Intel C++ Compiler Ver.~11.1 with the compile options \verb|-O3 -ip -ipo -xSSE4.2 -axSSE4.2 -openmp|.
A summary of the platforms is shown in Table~\ref{tbl_computers}.

Although we perform product runs of cavitation phenomena on both FX10 and SGI Altix ICE 8400EX,
we perform benchmark simulations only on FX10 at ITC, since
the benchmark results on ISSP have already been reported~\cite{mdnote}.

\subsection{Comparison between Flat MPI and Hybrid}
\label{sec_benchmark_conditions}

The conditions for our benchmark simulations are as follows~\cite{mdnote}.
All quantities are measured in LJ parameters, such as the particle diameter $\sigma$,
and the energy scale $\varepsilon$.
\begin{itemize}
\item The dimensions of each domain are $100 \times 100 \times 100$ and 
one domain is assigned to one MPI process (flat MPI) or one OpenMP thread (hybrid).
\item The number density is 0.5, \textit{i.e.}, each domain contains 500,000 particles.
\item Initial condition: Face-centered-cubic lattice.
\item Boundary condition: Periodic for all axes.
\item Integration scheme: Second-order symplectic integration.
\item Time step: 0.001.
\item Cutoff length: 2.5.
\item Cutoff scheme: Add constant and quadratic terms to potential as shown in Eq.~(\ref{eq_lj_cutoff}).
\item Initial velocity: The absolute value of the velocity of all particles is set to 0.9 and
 the directions of the velocities are given randomly.
\item After 150 steps, we measure the calculation (elapsed) time for the next 1000 steps.
\end{itemize}
Computational speeds are measured in the unit MUPS, which is millions of updates per second.
When a system with $N$ particles is simulated for $k$ steps in $t$ seconds, then
the number of MUPS is given by $ 10^{-6} Nk/t$.
The number of nodes is increased while keeping the same number of particles on each node (weak scaling).
We apply two paradigms, flat-MPI and the OpenMP/MPI hybrid for each size of simulation.
In the flat-MPI case, we assign 16 processes on a node (one MPI process to one CPU core).
In the hybrid case, we assign one MPI process containing 16 OpenMP threads on a node (one OpenMP thread to one CPU core).
We use identical executable file for the two models and  we assign domains to nodes
so that the same part of the system is assigned to the same CPU core for the flat-MPI and hybrid.
Therefore, the simulations are completely identical for the flat-MPI and hybrid runs including the communication patterns across nodes.

\begin{figure}
\begin{center}
\includegraphics[width=10cm]{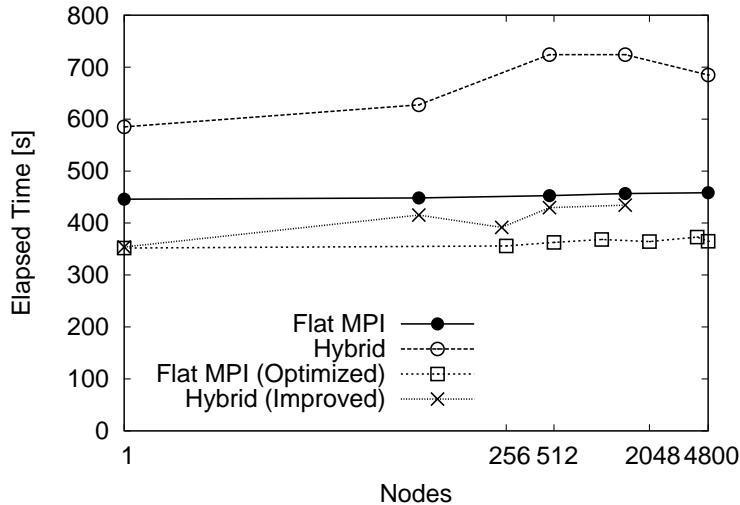}
\end{center}
\caption{Results of benchmark simulations for flat-MPI and hybrid runs (weak scaling).
Elapsed times for 1000 steps are shown. Although the elapsed times of flat-MPI are almost independent of the number of nodes,
the results of hybrid runs fluctuate. Additionally, computational speed of the flat-MPI is 
always higher than that of the hybrid.
The data of flat-MPI using the optimized code for FX10 is also shown which is denoted by ``Flat MPI (Optimized)".
The optimized code is about 25\% faster than the original code.
See Sec.~\ref{sec_fx10} for details.
The data of improved hybrid code is denoted by ``Hybrid (Improved)".
The computational speed is improved by about 65\% and it achieves almost same performance as that of the flat-mpi in a single node. See Sec.~\ref{sec_improve} for details.
}
\label{fig_sec}
\end{figure}

The results of the flat-MPI are summarized in Table~\ref{tbl_flatmpi} and those of the hybrid are summarized in Table~\ref{tbl_hybrid}.
The elapsed times of the runs are shown in Fig.~\ref{fig_sec}.
Since we perform weak scaling analysis, the elapsed time will be independent of the number of nodes
for perfect parallel efficiency. As shown in Fig.~\ref{fig_sec},
while the parallel efficiency of flat-MPI is almost perfect, that of the hybrid runs fluctuates.
The performance of flat-MPI is always better than that of the hybrid.
Since the computations are identical for the two methods, the difference 
originates from the parallel overheads of OpenMP.
There are several possible causes of the overheads.
One of them is the cost of thread management. The creation and destruction of threads are required for OpenMP.
Additionally, the number of synchronizations per step is larger than in flat-MPI.
Although only one global synchronization per step is required for a flat-MPI,
several intranode synchronizations are required for the hybrid.
Even for a perfect load balance, frequent synchronization may cause parallel overhead
due to noise from the system.
Note that the total memory required for the hybrid is less than that for flat-MPI, as expected.
For the largest run involving 4800 nodes, the hybrid run consumed 10 GB per node out of 32 GB
while flat-MPI consumed 14 GB.

\begin{table}[htb]
\begin{tabular}{rrrrr}
\hline
Nodes & Number of Particles & Elapsed Time [s] & Speed [MUPS] & Efficiency \\
\hline
1&  8,000,000 &      445.815 &17.9447& 1.0\\
72 & 576,000,000  &   448.275& 1284.93& 0.99\\
480 & 3,840,000,000 &  452.781& 8480.93& 0.98\\
1440 & 11,520,000,000& 456.641& 25227.7& 0.98\\
4800 &38,400,000,000 &458.399 &83769.9& 0.97\\
\hline
\end{tabular}
\caption{ Results of benchmark simulations of flat-MPI.
The number of nodes, number of particles, elapsed time [s], computational speed [MUPS], and parallel efficiency are listed.
The parallel efficiency is determined relative to the elapsed time for the single-node calculation.
}
\label{tbl_flatmpi}
\end{table}

\begin{table}[htb]
\begin{tabular}{rrrrr}
\hline
Nodes & Number of Particles & Elapsed Time [s] & Speed [MUPS] & Efficiency \\
\hline
1& 8,000,000&  585.160&13.6715&1.00\\
72 &  576000000 &  627.449 & 918.003 & 0.93\\
480 & 3840000000 & 724.139 & 5209.34 & 0.79 \\
1440 &11520000000 & 724.139 & 15908.5&  0.81 \\
4800&38,400,000,000&684.871&56068.9&0.85\\
\hline
\end{tabular}
\caption{ Same as Table~\ref{tbl_flatmpi} for hybrid runs.
}
\label{tbl_hybrid}
\end{table}

\subsection{Optimization for FX10}
\label{sec_fx10}

The results shown in the previous section are obtained using a code optimized for Intel Xeon architectures.
The performance of the largest job is 56.5 teraflops, which corresponds to an approximately 5\% execution efficiency.
To improve the efficiency and performance of simulations, we optimize the force calculation routine for FX10.
The algorithm optimized for FX10 is given as Algorithm~\ref{alg_simd}, where
$N$ is the total number of particles, $\textbf{q}[i]$ is the position of particle $i$,  $\textbf{p}[i]$ is the momentum,
and $dt$ is the time step.
The variables in bold denote three-dimensional vectors, \textit{e.g.}, $\textbf{q} = (q_x, q_y, q_z)$.
The constants $c_2$ and $c_0$ are coefficients introduced for the truncation in Eq.~(\ref{eq_lj_cutoff}).
We find that store operations involving indirect addressing is expensive for FX10.
Therefore, we avoid random storing by not using Newton's third law.
While the number of iterations required to calculate forces (the length of SortedList) is doubled without Newton's third law,
the computational efficiency increases more than twofold.
We also utilize SIMD operations. The variables with hats in Algorithm~\ref{alg_simd}
denote the packed variables.
A packed variable is stored in 128-bit registers and contains two double-precision 64-bit numbers.
Suppose there are two double-precision numbers $a$ and $b$ and a packed variable $\hat{c}$.
Packing $a$ and $b$ into $\hat{c}$ is denoted as
$\hat{c} \leftarrow \{a,b\}$ and 
unpacking $\hat{c}$ to $a$ and $b$
is denoted as $\{a,b\} \leftarrow  \hat{c}$.
To hide the latency by software pipelining, the inner loop is unrolled four times.
Since the two-body potential is truncated at the cutoff length $r_c$,
the force calculation involves checking whether or not the distance between a particle pair
is shorter than the cutoff length.
To perform the check with SIMD operations,
a new variable, $w=1-\lfloor r^2/r_c^2\rfloor$, is introduced, where $\lfloor x\rfloor$ denotes the largest integer not larger than $x$. 
The variable $w$ can be used as a mask for the truncation of the potential, since $w=1$ when the distance between the particle pair
is shorter than the cutoff length and $w=0$ otherwise.
Note that each distance between a particle pair in a pair list should be less than $\sqrt{2}r_c$
to use this mask. We construct a pair list so that this condition is always satisfied.

\begin{algorithm}[tb]
\caption{Code for calculating force optimized for FX10.}
\label{alg_simd}
\begin{algorithmic}[1]
\STATE $\hat{r}_c\leftarrow \{r_c,r_c\}$
\STATE $\hat{c}_2\leftarrow \{c_2,c_2\}$ 
\STATE  $\hat{u}\leftarrow \{1,1\}$
\STATE $\hat{dt}\leftarrow \{dt,dt\}$
\FOR{$i=1$ \TO $N-1$}
\STATE $\hat{\textbf{q}}_i\leftarrow \{\textbf{q}[i],\textbf{q}[i] \}$ 
\STATE $l\leftarrow\lfloor(\mathrm{KeyPointer}[i+1]-\mathrm{KeyPointer}[i])/4\rfloor$
\STATE $k_0\leftarrow \mathrm{KeyPointer}[i]$
\STATE  $\hat{\textbf{p}}_i\leftarrow \{\textbf{0},\textbf{0}\}$ 
\FOR{$k=0$ \TO $l-1$}
\STATE $k\leftarrow k_0+4k_1$
\STATE $j_a\leftarrow \mathrm{SortedList}[k]$ 
\STATE $j_b\leftarrow \mathrm{SortedList}[k+1]$
\STATE $j_c\leftarrow \mathrm{SortedList}[k+2]$
\STATE $j_d\leftarrow \mathrm{SortedList}[k+3]$
\STATE $\hat{\textbf{q}}_A\leftarrow \{\textbf{q}[j_a],\textbf{q}[j_b]\}$ 
\STATE $\hat{\textbf{q}}_B\leftarrow \{\textbf{q}[j_c],\textbf{q}[j_d]\}$ 
\STATE $\hat{\textbf{r}}_A\leftarrow\hat{\textbf{q}}_A-\hat{\textbf{q}}_i$ 
\STATE $\hat{\textbf{r}}_B\leftarrow\hat{\textbf{q}}_B-\hat{\textbf{q}}_i$ 
\STATE $\hat{r}_A^2\leftarrow |\hat{\textbf{r}_A}|^2$ 
\STATE $\hat{r}_B^2\leftarrow |\hat{\textbf{r}_B}|^2$ 
\STATE $\hat{w}_A\leftarrow \hat{u}-\lfloor\hat{r}_A^2/\hat{r}_c^2\rfloor$ 
\STATE $\hat{w}_B\leftarrow \hat{u}-\lfloor\hat{r}_B^2/\hat{r}_c^2\rfloor$ 
\STATE $\hat{f}_A\leftarrow [(24\hat{r}_A^6-48)/\hat{r}_A^{14}+8\hat{c}_2] \hat{dt}  \hat{w}_A$ 
\STATE $\hat{f}_B\leftarrow [(24\hat{r}_B^6-48)/\hat{r}_B^{14}+8\hat{c}_2] \hat{dt}  \hat{w}_B$ 
\STATE $\hat{\textbf{p}}_i\leftarrow \hat{\textbf{p}}_i + \hat{f}_A\hat{\textbf{r}}_A+\hat{f}_B\hat{\textbf{r}}_B$ 
\ENDFOR
\STATE $\{\textbf{p}_1,\textbf{p}_2\}\leftarrow \hat{\textbf{p}}_i$ 
\STATE $\textbf{p}[i]\leftarrow \textbf{p}[i]+\textbf{p}_1+\textbf{p}_2$ 
\FOR{$k = $ KeyPointer[$i$] $+4l$ \TO KeyPointer[$i+1$]-1}
\STATE $j\leftarrow \mathrm{SortedList}[k]$ 
\STATE $\textbf{r}\leftarrow \textbf{q}[j]-\textbf{q}[i]$
\STATE $r^2\leftarrow |\textbf{r}|^2$ 
\IF{$r^2 < r_c^2$}
\STATE $f\leftarrow [(24r^6-48)/r^{14}+8c_2] dt$
\STATE  $\textbf{p}[j]\leftarrow \textbf{p}[j]+f \textbf{r}$
\ENDIF
\ENDFOR
\ENDFOR
\end{algorithmic}
\end{algorithm}

\begin{figure}[htb]
\begin{center}
\includegraphics[width=10cm]{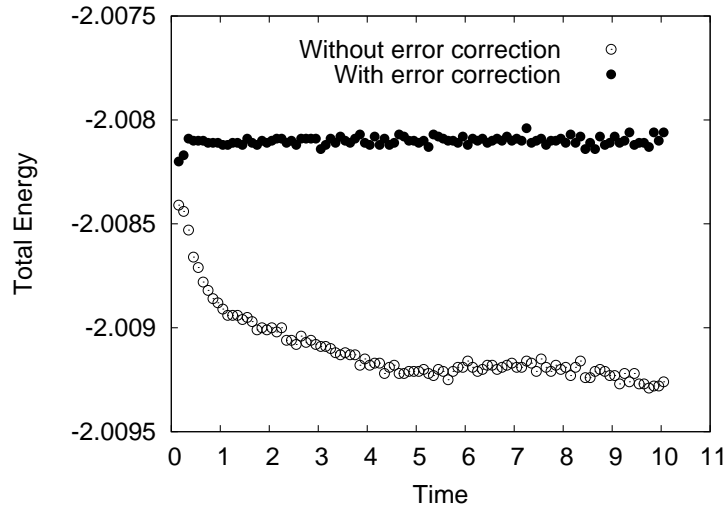}
\end{center}
\caption{
Effect of error correction on reciprocal approximation.
The number of particles is 4000. Computations are performed on a single core.
The open circles denote the results without the error correction
and the filled circles denote those with the error correction.
The energy conservation with the error correction is sufficient.
}
\label{fig_err}
\end{figure}

The force calculation of the LJ interaction involves at least one division for each particle pair.
However, the floating-point divide instruction (fdivd) in FX10 cannot be specified as a SIMD instruction.
Therefore, we use the floating-point reciprocal approximation instruction (frcpad) instead of the divide instruction.
The reciprocal approximation instruction is faster than the divide instruction at the expense of precision.
The rounding error due to the approximation is up to $1/256$, \textit{i.e.},
\begin{equation}
\left|  \frac{\mathrm{frcpad}(x) - 1/x }{1/x} \right| < \frac{1}{256},
\end{equation}
which is  insufficient for MD simulations.
Therefore, we improve the precision by the following error correction.
Suppose $r$ is the particle distance.
The force calculation of 6-12 LJ interactions requires the values of $r^{-8}$ and $r^{-14}$.
Consider the calculation of $r^{-8}$ from $r^2$.
Let $\tilde{r}^{-2}$ be a reciprocal approximation of $r^2$, \textit{i.e.}, $\tilde{r}^{-2} = \mathrm{frcpad}(r^2)$.
The rounding error due to the approximation can be expressed as
\begin{equation}
\tilde{r}^{-2} r^2 =1 + \varepsilon, \qquad (|\varepsilon| < 1/256)
\end{equation}
with error $\varepsilon$.
Then the required value $r^{-8}$ is expressed as
\begin{eqnarray}
r^{-8} &=& \frac{(\tilde{r}^{-2})^4}{(1+\varepsilon)^4} \\
&=& (\tilde{r}^{-2})^4 (1- 4 \varepsilon + O(\varepsilon^2)) \\
&=&   (\tilde{r}^{-2})^4 (5 - 4(1+ \varepsilon) + O(\varepsilon^2)) \\
&=& (\tilde{r}^{-2})^4 (5 - 4 \tilde{r}^{-2} r^2 + O(\varepsilon^2)).
\end{eqnarray}
Therefore, the error of the corrected value $(\tilde{r}^{-2})^4 (5 - 4 \tilde{r}^{-2} r^2)$ relative to $r^{-8}$
is $O(\varepsilon^2)$.
Similarly, the correction coefficient for $r^{-14}$ is obtained to be $(8 - 7 \tilde{r}^{-2} r^2)$.
The effect of the error correction is shown in Fig.~\ref{fig_err}.
One can see that the conservation of the total energy is improved by the error correction.

The performance of the optimized code is summarized in Table~\ref{tbl_flatmpi_suzuki}
and shown in Fig.~\ref{fig_sec} as the data denoted by Flat MPI (Optimized).
The increase in speed due to optimization is about 25\% compared with the original code with flat MPI.
Although the performance also increases to 193 teraflops, which corresponds to a 17.0\% execution efficiency,
one has to note that it contains redundant computations owing to the nonuse of Newton's third law.

\begin{table}[htb]
\begin{tabular}{rrrrr}
\hline
Nodes & Number of Particles & Elapsed Time [s] & Speed [MUPS] & Efficiency \\
\hline
1   & 8,000,000     &351.821 &22.7389 &1.00\\
256  & 2,048,000,000 & 355.829 &5755.57& 0.99\\
512  & 4,096,000,000  &362.775 &11290.7 &0.97\\
1024 & 8,192,000,000  &368.275& 22244.2 &0.96\\
2048 & 16,384,000,000& 364.272 &44977.4& 0.97\\
4096 & 32,768,000,000 &373.007 &87848.3 &0.94\\
4800 & 38,400,000,000 &364.920 &105229.0 &0.96\\
\hline
\end{tabular}
\caption{Same as Tables~\ref{tbl_flatmpi} and \ref{tbl_hybrid} for 
the flat-MPI code optimized for FX10.
}
\label{tbl_flatmpi_suzuki}
\end{table}

\subsection{Improvements in Hybrid Code}
\label{sec_improve}

The simulations presented in the manuscript were performed in the large queue
as ``Large-scale HPC Challenge" Project at Information Technology Center of the University of Tokyo.
After the project, we found the reason why hybrid runs were considerably slower than the flat-mpi runs
at FX10 with the help of FUJITSU.
The cause is \verb|std::vector| in the Standard Template Library.
We use \verb|std::vector| in the routines for searching interacting particle-pairs,
and it becomes slow when it runs in multi-thread environment.
We use  \verb|std::vector| as a temporary buffer as follows.
\begin{verbatim}
void
SomeClass::SomeFunction(void){
   std::vector<int> v;
  // some procedure with v.
}
\end{verbatim}
The function \verb|SomeFnction| can be executed simultaneously by each thread,
and exclusive access control is not necessary.
Therefore, we change the code so that the \verb|std::vector| becomes 
thread-local variable as,
\begin{verbatim}
void
SomeClass::SomeFunction(void){
   static __thread std::vector<int> v;
   v.clear();
  // some procedure with v.
}
\end{verbatim}
We performed benchmarks in order to investigate the effect of the above modification.
We used Fujitsu C/C++ Compiler Ver.~1.2.1 with the compile options \verb|-Kfast -Ksimd=2 -Knoparallel -Kopenmp -Xg|.
We apply the optimization described in Sec.~\ref{sec_fx10} for this run.
The conditions for the benchmarks are same as that described in Sec.~\ref{sec_benchmark_conditions}.
The benchmark results are summarized in Table~\ref{tbl_hybrid_improved} and shown in Fig.~\ref{fig_sec}.
The computational speed of hybrid run at single-node calculation is much improved
compared with the results in Table.~\ref{tbl_hybrid}, and it achieves almost same performance
as that of the flat-mpi. However, the computational efficiency decreases 
as a number of nodes increases, while the results of flat-mpi show almost perfect parallel efficiency.

\begin{table}[htb]
\begin{tabular}{rrrrr}
\hline
Nodes & Number of Particles & Elapsed Time [s] & Speed [MUPS] & Efficiency \\
\hline
1 &8,000,000 & 353.977 & 22.6003& 1.00 \\
72 &576,000,000 & 415.281 &1387.01 & 0.85 \\
240 &1,920,000,000 & 391.666 &4902.13 &   0.90 \\
480 &3,840,000,000 & 429.698 &8936.51 &  0.82 \\
1440 &11,520,000,000 & 434.544 &26510.5&  0.81\\
\hline
\end{tabular}
\caption{Results of benchmark simulations of hybrid runs with the improved code.
The results without the improvements are summarized in Table~\ref{tbl_hybrid}.
}
\label{tbl_hybrid_improved}
\end{table}

\subsection{Strong Scaling Analyses}
\label{sec_strong}

In order to investigate the difference of performance between processes and threads, 
we perform strong scaling analyses, \textit{i.e.}, we increase a number of
processes or threads keeping a total number of particles.
We perform benchmarks on a single node, since
amount of inter-node communication as well as the pattern of communication
are equivalent between the flat-MPI and the hybrid in the pseudo-flat-MPI approach,
and the difference between them only appears in a node.
Hereinafter, we refer to the hybrid as the flat-OpenMP, since we perform hybrid benchmarks
with only one MPI process.
Let $T_n$ be elapsed time using $n$ processes/threads. We define the parallel efficiency $\alpha$
as,
\begin{equation}
\alpha = \frac{T_1}{n T_n}.
\end{equation}

\begin{figure}[htb]
\begin{center}
\includegraphics[width=10cm]{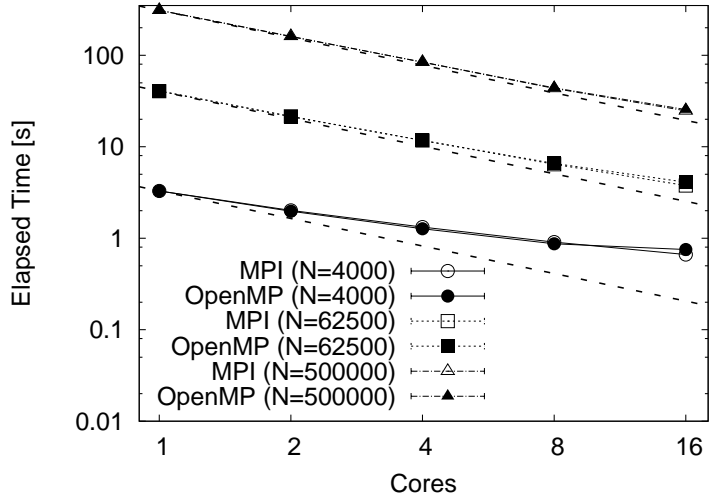}
\end{center}
\caption{Benchmark results for strong scaling analyses in a single node.
Results of flat-mpi and flat-openmp are shown for systems with different total number of particles.
The results of flat-MPI and flat-OpenMP are denoted by the open and filled symbols.
The dashed lines denote the ideal elapsed times with perfect efficiency.
The fluctuations in execution time are smaller than the size of the symbols.
}
\label{fig_strong}
\end{figure}

We investigate three sizes, 4000, 62500, and 500000 particles and
increase a number of processes or threads from 1 to 16 for each system size.
In order to evaluate fluctuation in execution time, we perform three identical runs for each system size
and a number of processes or threads, and compute the standard deviations of the elapsed times.
The results are summarized in Tables~\ref{tbl_strong_mpi} and \ref{tbl_strong_openmp}
and shown in Fig.~\ref{fig_strong}.
While the computational performance (MUPS) are similar between $N=62,500$ and $N=500,000$ cases,
it drops drastically when $N=4,000$. 
The parallel efficiency relative to a single process/thread calculation decreases drastically
when a number of particles on each process/thread is less than one thousand.
While the calculations using 16 threads (flat-OpenMP) are slightly but significantly slower than those using
16 processes, there are no apparent difference between the flat-MPI and the flat-OpenMP.
We also find that the fluctuations in the execution time of the flat-OpenMP are small
and comparable to those of the flat-MPI.
Therefore, the reason why the results of hybrid-runs involving inter-node
communication fluctuate, as observed in Fig.~\ref{fig_sec}, is still an open question.

\begin{table}[htb]
\begin{tabular}{rrrrr}
\hline
Processes & Particles & Elapsed Time [s] & Speed [MUPS] & Efficiency \\
\hline
1 & 4,000 & 3.290(2) & 1.2158(8) & 1.00 \\
2 & 4,000 & 2.0179(8) & 1.9823(8) & 0.82 \\
4 & 4,000 & 1.325(2) & 3.019(5) & 0.62 \\
8 & 4,000 & 0.906(3) & 4.41(2) & 0.45 \\
16 & 4,000 & 0.664(6) & 6.03(6) & 0.31 \\
\hline
1 & 62,500 & 40.549(8) & 1.5413(3) & 1.00 \\
2 & 62,500 & 21.50(1) & 2.907(2) & 0.94 \\
4 & 62,500 & 11.73(1) & 5.327(5) & 0.86 \\
8 & 62,500 & 6.442(1) & 9.701(2) & 0.79 \\
16 & 62,500 & 3.763(3) & 16.61(1) & 0.67 \\
\hline
1 & 500,000 & 310.9(1) & 1.6083(6) & 1.00 \\
2 & 500,000 & 161(1) & 3.10(2) & 0.96 \\
4 & 500,000 & 84.1(3) & 5.94(2) & 0.92 \\
8 & 500,000 & 43.46(2) & 11.505(4) & 0.89 \\
16 & 500,000 & 24.687(7) & 20.254(6) & 0.79 \\
\hline
\end{tabular}
\caption{Results of benchmark simulations of flat-MPI. The number of cores, number of particles, elapsed time [s], computational-   speed [MUPS], and parallel efficiency are listed. The parallel efficiency is determined relative to the elapsed time for the single-core calculation.
}
\label{tbl_strong_mpi}
\end{table}

\begin{table}[htb]
\begin{tabular}{rrrrr}
\hline
Threads & Particles & Elapsed Time [s] & Speed [MUPS] & Efficiency \\
\hline
1 & 4000 & 3.286(3) & 1.217(1) & 1.00 \\
2 & 4000 & 1.974(2) & 2.026(2) & 0.83 \\
4 & 4000 & 1.269(6) & 3.15(2) & 0.65 \\
8 & 4000 & 0.869(5) & 4.60(3) & 0.47 \\
16 & 4000 & 0.75(1) & 5.31(9) & 0.27 \\
\hline
1 & 62500 & 40.913(2) & 1.52765(9) & 1.00 \\
2 & 62500 & 21.320(7) & 2.931(1) & 0.96 \\
4 & 62500 & 11.72(3) & 5.33(1) & 0.87 \\
8 & 62500 & 6.577(7) & 9.503(9) & 0.78 \\
16 & 62500 & 4.103(9) & 15.23(3) & 0.62 \\
\hline
1 & 500000 & 310.100(2) & 1.612383(8) & 1.00 \\
2 & 500000 & 159.898(9) & 3.1270(2) & 0.97 \\
4 & 500000 & 84.3(2) & 5.93(1) & 0.92 \\
8 & 500000 & 44.0(2) & 11.36(5) & 0.88 \\
16 & 500000 & 25.53(10) & 19.59(7) & 0.76 \\
\hline
\end{tabular}
\caption{Same as Table~\ref{tbl_strong_mpi} for the flat-OpenMP.
}
\label{tbl_strong_openmp}
\end{table}

\section{Multibubble Nuclei} \label{sec_cavitation}

\begin{figure}[htb]
\begin{center}
\includegraphics[width=8cm]{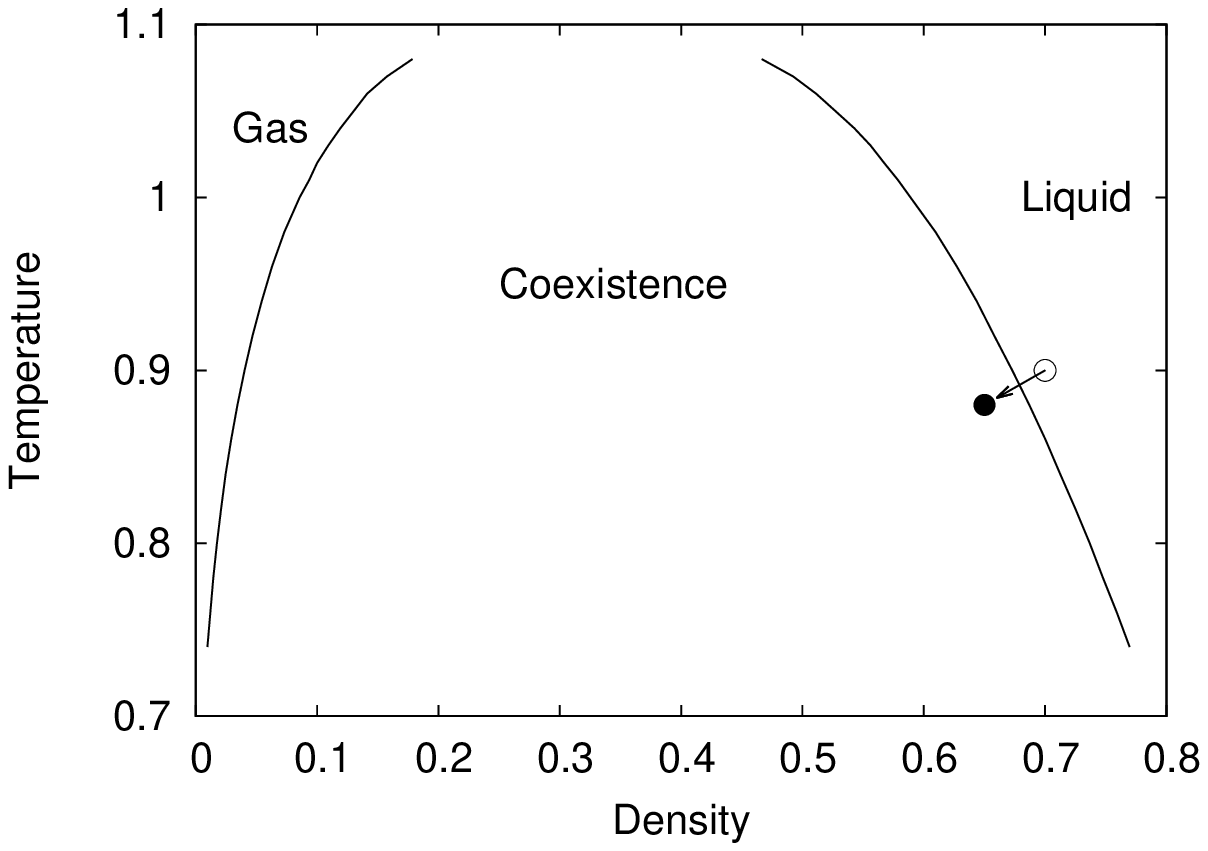}
\end{center}
\caption{ Phase diagram of the system.
The solid lines denote the gas-liquid coexistence densities.
The open circle denotes the initial condition of the system.
After the expansion, the state of the system moves to the point denoted by the solid circle.
}
\label{fig_phase}
\end{figure}

The study of multibubble nuclei is challenging,
since it is a typical multiscale and multiphysics phenomenon that involves
phase transitions at the microscopic scale and interbubble interactions at the meso- or macroscopic scale.
From the viewpoint of computational simulations, phase transitions make simulations difficult
since they involve the creation, movement, and annihilation of phase interfaces.
If one simulates bubble nuclei using MD, then there is no need to consider
phase transitions and phase interfaces. Both of these occur spontaneously in the simulation
as a result of the dynamics of particles. However, if one wants to reproduce multibubble nuclei by MD, 
then a huge number of particles are required; therefore, a large-scale simulation is necessary.
Additionally, gas-liquid phase transitions involve large change in density and highly inhomogeneous density
could be cause serious load imbalance in MD simulations.

There are several reports on extremely large-scale MD simulations.
A team from Lawrence Livermore National Laboratory and IBM studied
the Kelvin-Helmholtz instability by MD involving 62.5 billion particles on BlueGene/L, for which they
won the 2007 Gordon Bell Prize~\cite{Glosli2007}.
In 2008, Germann and Kadau demonstrated one-trillion-atom MD on BlueGene/L using single precision~\cite{Germann2008}.
There are several reports on extremely large-scale MD simulations.
A team from Lawrence Livermore National Laboratory and IBM studied
the Kelvin-Helmholtz instability by MD involving 62.5 billion particles on BlueGene/L, for which they
won the 2007 Gordon Bell Prize~\cite{Glosli2007}.
In 2008, Germann and Kadau demonstrated one-trillion-atom MD on BlueGene/L using single precision~\cite{Germann2008}.
There are also huge simulations involving phase transitions.
Streitz et al.~performed MD simulation involving 32,768,000 on 65576 processors
and investigated the finite-size effect on solidifications~\cite{Streitz2006}.
Shimokawabe et al.~demonstrated simulations of solidification with phase-field method
using 5,000 GPUs along with 16,000 CPU cores~\cite{Shimokawabe2011}.
Despite the above challenges, to our best knowledge, there have not been studies of large MD simulations
involving gas-liquid phase transitions.

In this section, we describe our large-scale MD simulations of multibubble nuclei.
There are two types of processes involving bubble nuclei; boiling and cavitation.
Boiling occurs when a liquid is heated under a constant-pressure condition and
cavitation occurs when the pressure of the liquid is reduced.
We study cavitation phenomena since a simulation of cavitation requires only a thermostat, while
that of boiling requires both a thermostat and a barostat.

\subsection{Cavitation Process}

\begin{table}[htbp]
\begin{tabular}{ccrrrrrc}
\hline
Site & Scheme  & Processes & Threads & System Size &Particles&Total  Steps & Elapsed Time [h] \\ \hline
ISSP & Flat-MPI & 1,024 &  1,024 &  $320\times 320 \times 320$  &22,937,600   &  690,000 & 10 \\ \hline
ITC & Hybrid& 4,800      & 76,800 & $1,440 \times 1,200 \times 1,200$   &1,449,776,020 & 178,000 & 15 \\ \hline
\end{tabular}
\caption{ Details of conditions. The parallelization scheme, the numbers of processes and threads used for the jobs,
the dimensions of the system, the total number of particles, the total number of steps, and the elapsed time are shown
for each site.
} \label{tbl_conditions}
\end{table}

We first maintain the system at a temperature $T=0.9$ with a density $\rho=0.7$ using a thermostat.
This corresponds to the pure-liquid phase (see Fig.~\ref{fig_phase} for the phase diagram.).
We use the Nos\'e--Hoover thermostat to control temperature~\cite{NoseHoover}.
Then we expand the system. The expansion is performed by
changing the length scale of the system from $l$ to $\alpha l$.
The rescaling factor $\alpha$ is chosen to be larger than unity.
With this change, the position of each particle also changes from $(q_x, q_y, q_z)$
to $(\alpha q_x, \alpha q_y, \alpha q_z) $ while keeping the momenta.
This procedure models a uniform and adiabatic expansion.
We chose the rescaling factor to be $1.025$. Then the density changes from $0.7$ to $0.65$,
which is in the gas-liquid coexistence region in the phase diagram; therefore, bubbles start to appear.
After expansion, we turn off the thermostat so that
it does not affect the physical processes of the bubble nuclei.
The integration scheme used for the isothermal time evolution is the second-order reversible system propagator algorithm (RESPA)~\cite{Tuckerman1992}, and the second order symplectic method is used for the microcanonical simulation with a time step $\Delta t =0.005$.
We perform a small job at ISSP and a large job at ITC.
Hereinafter, we denote the smaller and larger jobs by ISSP and ITC, respectively. 
The conditions of the jobs are summarized in Table~\ref{tbl_conditions}.

\subsection{Bubble Identification}

To identify bubbles, we divide the system into small subcells with a length of $3.0$
and determine the local density for each subcell~\cite{wtime}. 
The densities of the gas and liquid coexisting in this system at $T=0.9$ are estimated to be $0.0402(2)$ and 
$0.6730(2)$, respectively~\cite{phasediagram}. 
Therefore, we define a subcell to be in the {\it gas state} when its density is less than $0.2$.
We have confirmed that the results do not change for other threshold values such as $0.3$.
We consider that neighboring {\it gas state} cells are in the same cluster and
identify the bubbles using the site-percolation criterion on the simple cubic lattice.
This process is called clustering, and the parallelization of clustering is quite expensive
since it involves global communication similar to that of the fast fourier transform.
Therefore, we gather all the data of local densities to the root node
and perform clustering with serial calculation. To reduce memory consumption,
we count the number of particles in each subcell and store it as an integer (data type \verb|unsigned char| in C language)
instead of storing the local density with floating-point variables.
Since the volume of each subcell is 27 and the local density hardly exceeds 1.0,
the number of particles in a subcell is at most 27, which can be stored in one byte without losing any information.
A snapshot of the data is 1.4 MB at ISSP and 76.8 MB at ITC.
To perform clustering, more work memory is required in addition to the data of the local densities.

\subsection{Simulation Results}

\begin{figure}[htb]
\begin{center}
\includegraphics[width=8cm]{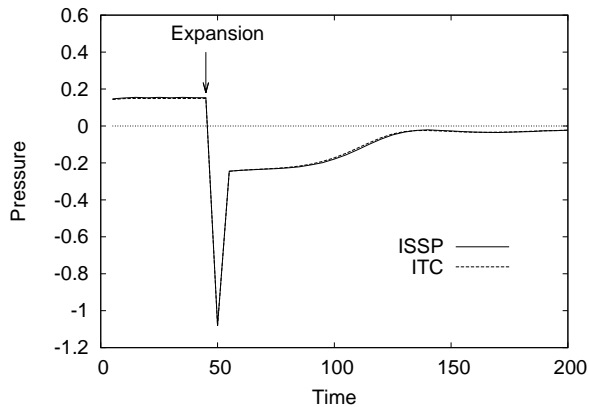}
\end{center}
\caption{Time evolutions of pressure.
First the system is maintained in the pure liquid phase. At $t=50$, the system expands
and the pressure becomes negative. As bubbles appear and grow, the pressure increases after the expansion.
The behaviors of the two jobs are almost identical.
}\label{fig_pressure}
\end{figure}

\begin{figure}[htb]
\begin{center}
\includegraphics[width=0.24\linewidth]{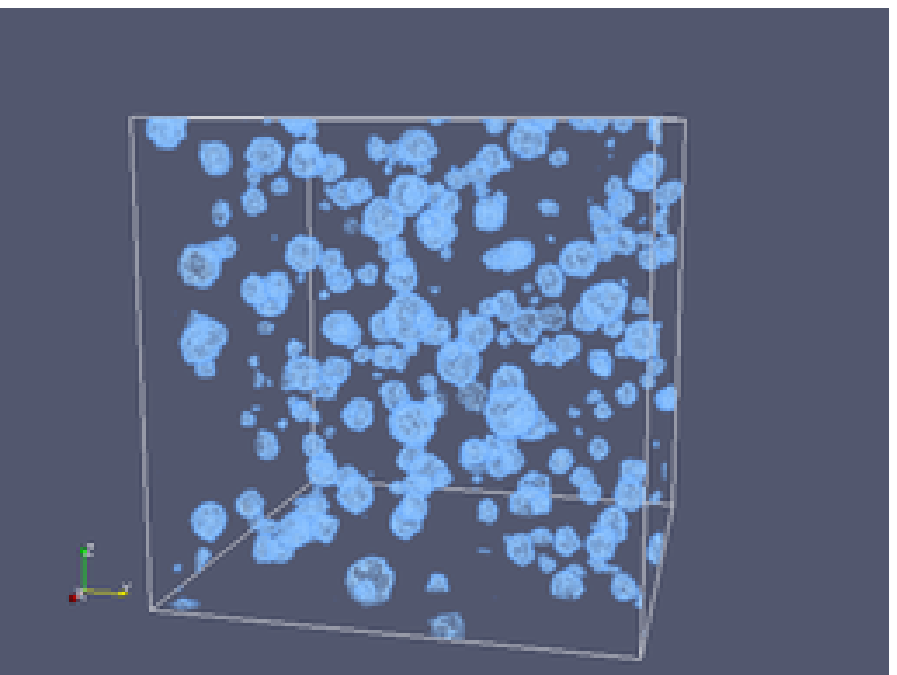}
\includegraphics[width=0.24\linewidth]{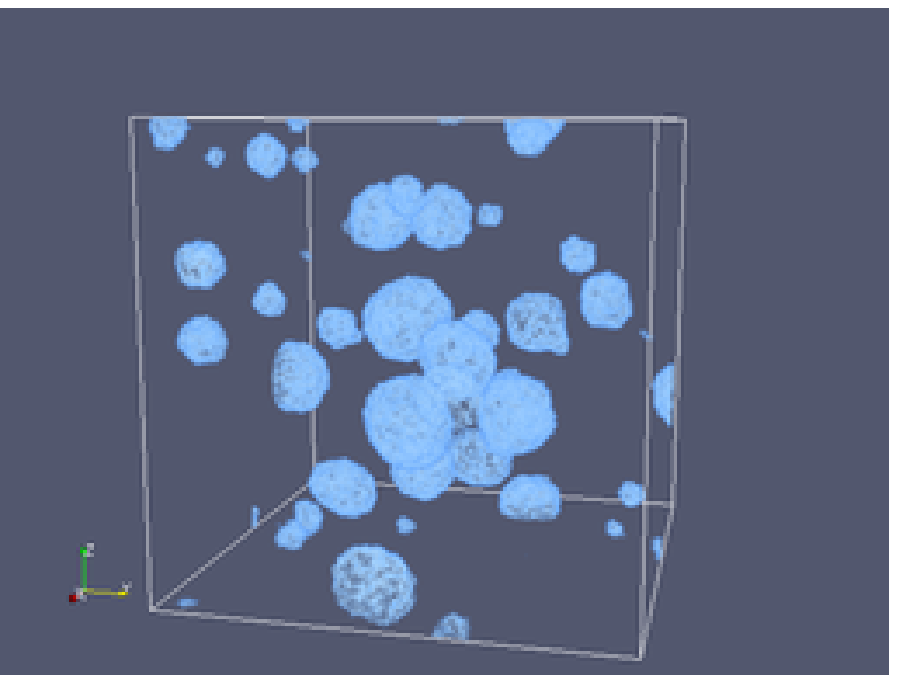}
\includegraphics[width=0.24\linewidth]{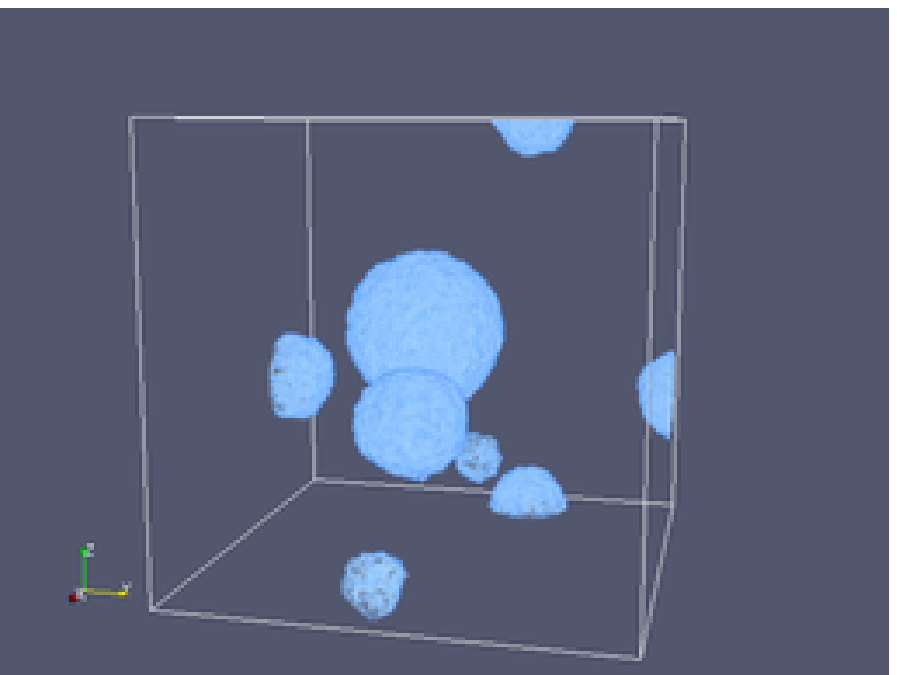}
\includegraphics[width=0.24\linewidth]{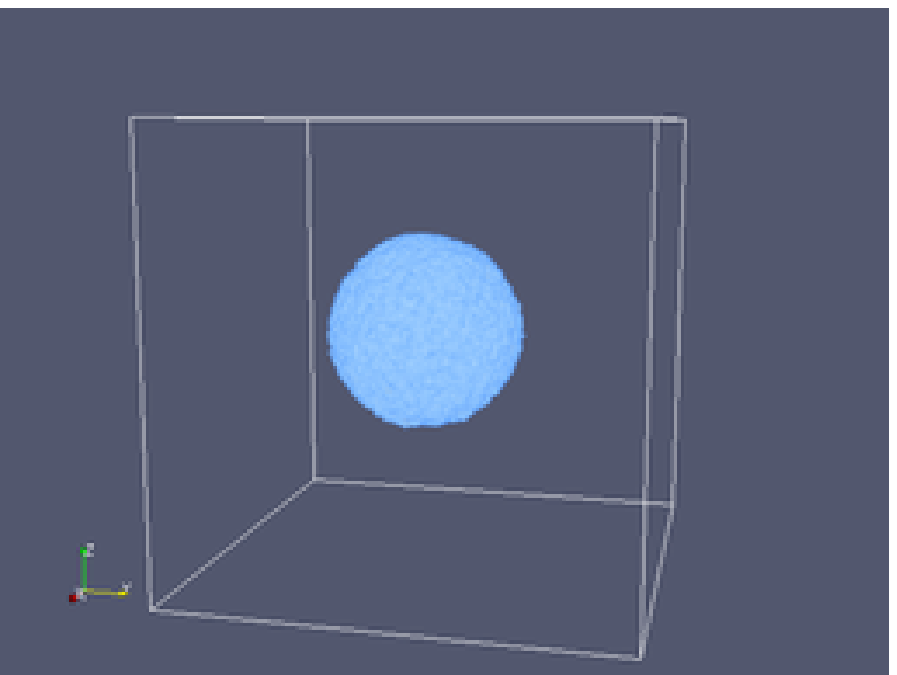}
\end{center}
\caption{
(Color online)  Snapshots of bubbles obtained during runs at ISSP. The direction of time evolution is from left to right.
In the early stages, many bubbles appear and grow independently.
Afterward, the bubbles start to interact and
Ostwald-like ripening, where larger bubbles become larger and smaller bubbles disappear, is observed.
Finally, one large bubble remains in the system.
These pictures were produced by ParaView~\cite{paraview}.
}
\label{fig_snapshots_issp}
\end{figure}

\begin{figure}[htb]
\begin{center}
\includegraphics[width=0.24\linewidth]{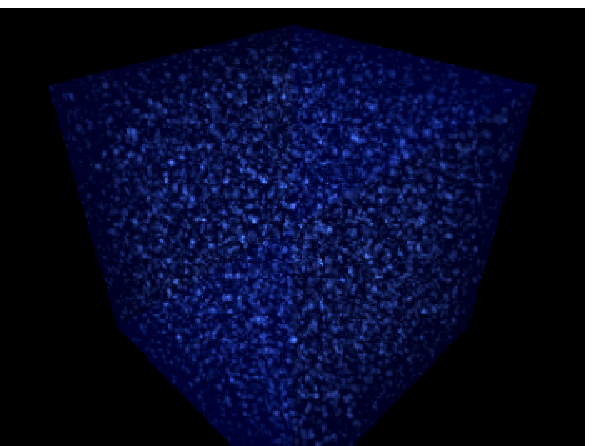}
\includegraphics[width=0.24\linewidth]{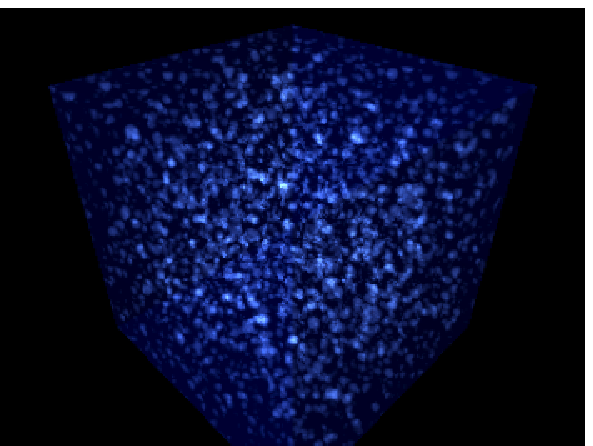}
\includegraphics[width=0.24\linewidth]{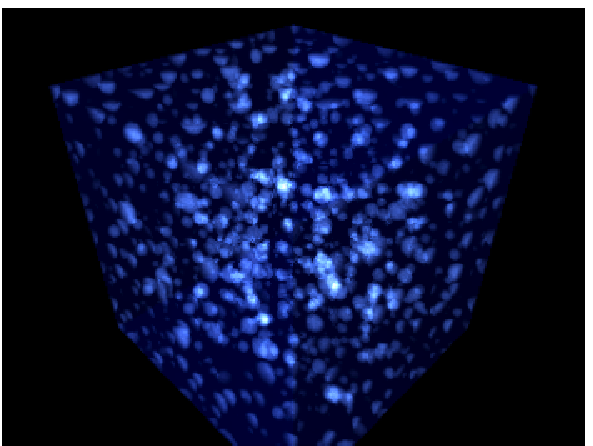}
\includegraphics[width=0.24\linewidth]{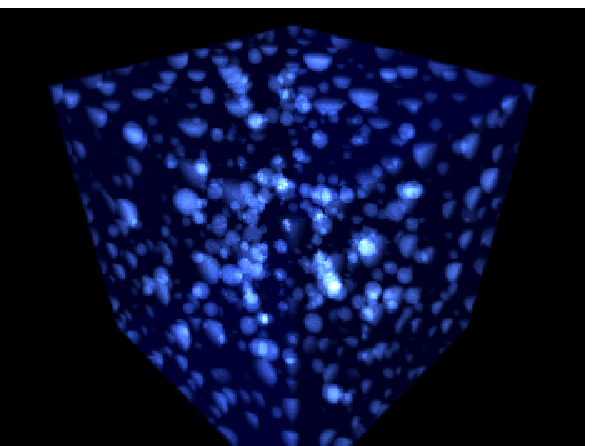}
\end{center}
\caption{
(Color online) Snapshots of bubbles obtained during runs at ITC. The direction of time evolution is from left to right.
Multibubble nuclei and the Ostwald-like ripening are observed.
These pictures were produced by POV-Ray~\cite{povray}.
}
\label{fig_snapshots_itc}
\end{figure}

We perform the job with flat-MPI at ISSP, whereas
we perform the job with the hybrid at ITC in order not to use too much memory.
The time evolution of pressure in each case is shown in Fig.~\ref{fig_pressure}.
Although the pressure is positive when the system is in the pure liquid phase,
it suddenly becomes negative after the expansion and increases as bubbles appear and grow.
Snapshots obtained during the runs are shown in Figs.~\ref{fig_snapshots_issp} and \ref{fig_snapshots_itc}.
In the job at ISSP, Ostwald-like ripening is observed after multibubble nuclei,
and finally only one bubble remains in the system.
Multibubble nuclei and Ostwald-like ripening are also observed at ITC.

\begin{figure}[htb]
\begin{center}
\includegraphics[width=8cm]{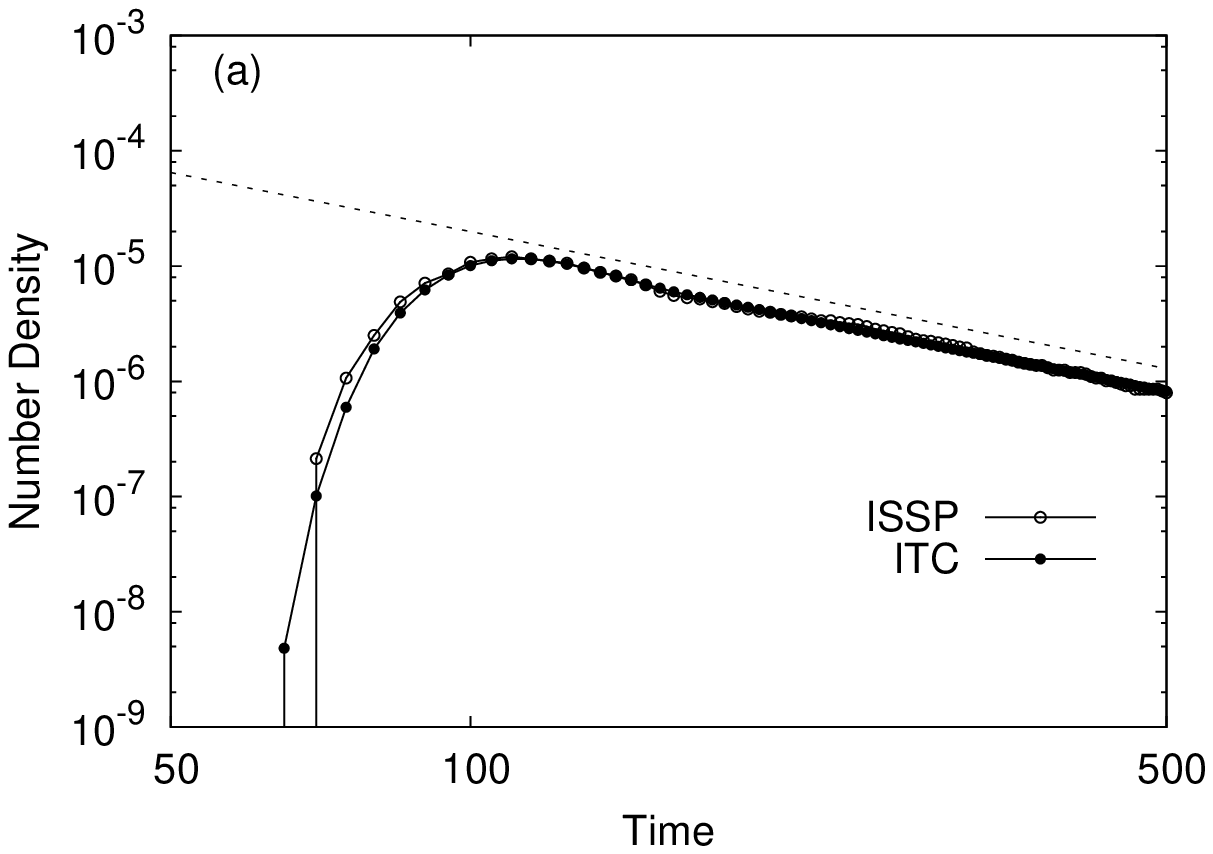}
\includegraphics[width=8cm]{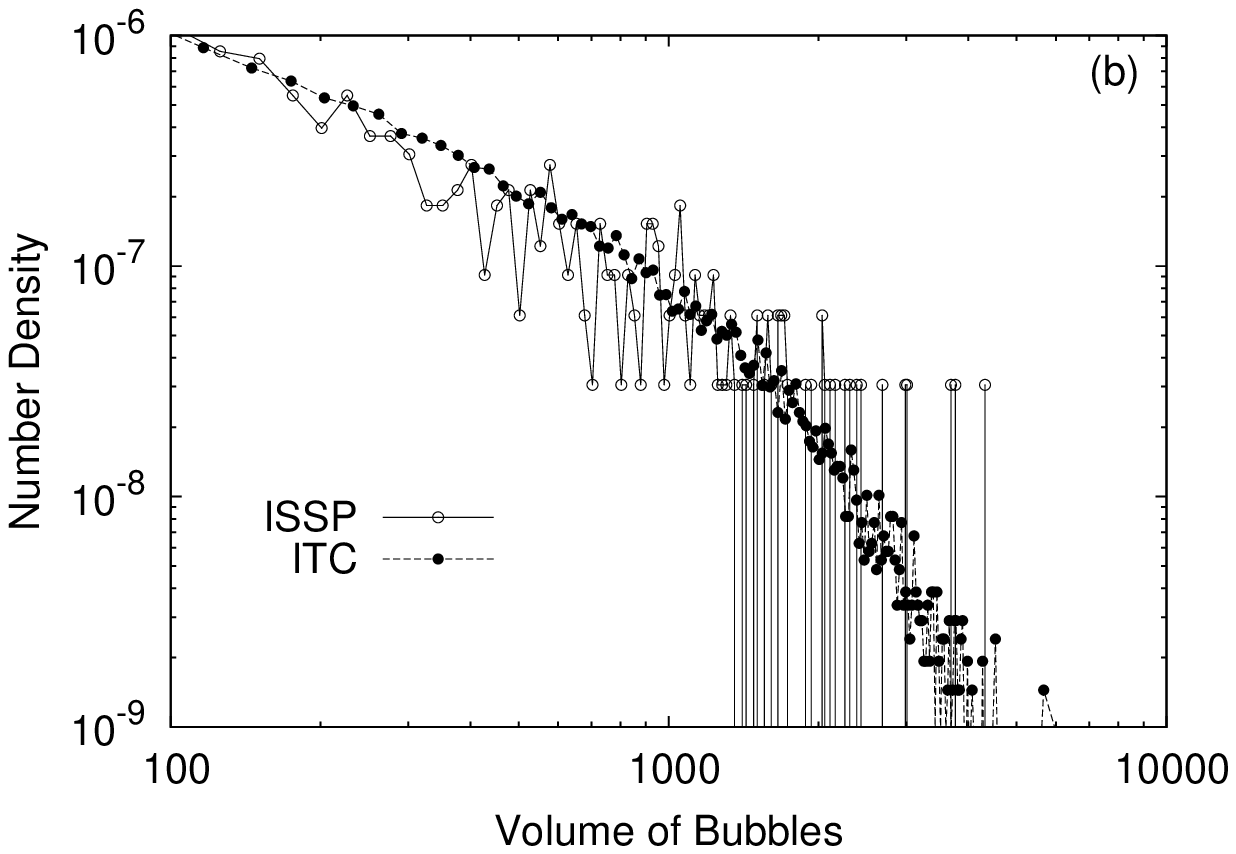}
\end{center}
\caption{(a) Time evolution of the total number of bubbles. 
The decimal logarithms are taken for both axes.
The dashed line with a slope of -1.7 is a guide to the eyes.
(b) Size distributions of bubbles at the 10,000th step after expansion.
The decimal logarithms are taken for both axes.
}
\label{fig_bubble}
\end{figure}

The time evolutions of the total number of bubbles are shown in Fig.~\ref{fig_bubble}~(a).
After the rapid increase in the number of bubbles,
the number decreases slowly owing to interactions between bubbles, \textit{i.e.},
small bubbles are eliminated owing to the growth of larger bubbles.
The power-law decay of the number of bubbles can be observed.
The distributions of the bubble sizes at the 10,000th step after expansion
are shown in Fig.~\ref{fig_bubble}~(b).
Although the results of both simulations show similar distributions,
one can see that the distribution for the jobs at ISSP appears to be insufficient for further analysis.

\subsection{Load Imbalance} \label{sec_loadimbalance}

\begin{figure}[tb]
\begin{center}
\includegraphics[width=12cm]{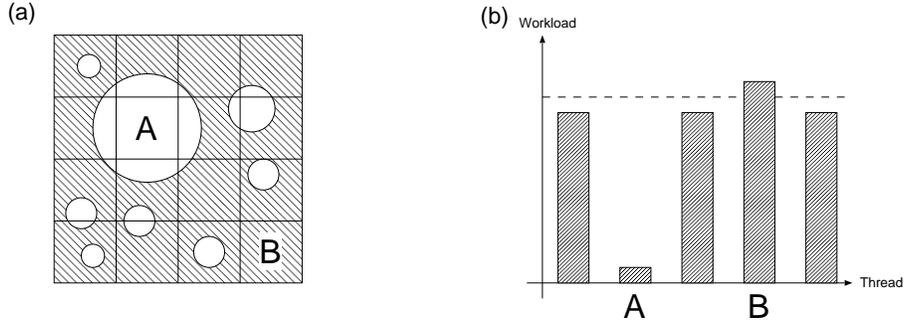}
\end{center}
\caption{
Schematic illustrations of load imbalance.
(a) Typical configuration of simulation of cavitation. There are many bubbles.
The domain denoted by \textit{A} is filled with pure gas, and the domain denoted by \textit{B}
is filled with pure liquid, \textit{i.e.}, it does not contain any bubbles.
(b) Workloads of threads are proportional to the densities of the assigned domains.
The dashed line denotes the average density of the system, which is the workload
of domains for a perfect load balance.
The workload of thread \textit{B} is about 20 times larger than that of thread \textit{A}.
However, it is only 5\% larger than the average workload.
}
\label{fig_loadimbalance}
\end{figure}

In the final part of this section, we discuss the load imbalance of our simulations.
To check the validity of the pair list, we synchronize all threads at every time step.
Therefore, the slowest thread determines the overall performance since the other threads must wait
for the slowest thread due to the synchronization.
The computational workload of a domain is roughly proportional to the number of particles it contains.
Since the volumes of domains are identical, the workload is proportional to the number density.
Since a liquid is almost incompressible, the largest density is determined by the coexisting density of the liquid 
at the temperature of the system. 
In our simulation, the temperature becomes 0.88 after the expansion.
The densities of the coexisting gas $\rho_\mathrm{g}$ and liquid $\rho_\mathrm{l}$ at this temperature are
$\rho_\mathrm{g} = 0.0344$ and $\rho_\mathrm{l} = 0.687$, respectively~\cite{phasediagram}.
Therefore, the computational time required for one step of the slowest process is proportional to $\rho_\mathrm{l}$,
which is 20 times larger than that of the fastest thread.
If a perfect load balance is achieved, all threads have identical numbers of particles,
and the densities in the domains become the same as the total number density $\rho_\mathrm{ave}$.
In our simulation, the average density after expansion is $\rho_\mathrm{ave} =0.65$.
Therefore, the computational time of the simulation without load balancing
is proportional to $\rho_\mathrm{l}$ and that with a perfect load balance is proportional to $\rho_\mathrm{ave} =0.65$.
The reduction in speed due to the load imbalance is given by the ratio of the average workload
to the heaviest workload, which is $1 - \rho_\mathrm{ave}/\rho_\mathrm{l} \sim 0.05$,
\textit{i.e.}, only 5\% (see Fig.~\ref{fig_loadimbalance}). 
This means that the load imbalance is a minor problem in studies of cavitation phenomena.
If one wants to study condensation phenomena, which involve droplet nuclei,
then the problem of load imbalance becomes serious.
Since the average density is close to that of the gas while the slowest thread is determined by the density of the liquid,
the simulation performance can be 20 times worse than that of the perfectly load-balanced simulation.

\section{Summary and Discussion} \label{sec_summary}

We have developed a pseudo-flat MPI MD code
and performed benchmark and product runs utilizing up to 4800 nodes, and 76,800 cores.
Benchmark simulations contained up to 38.4 billion particles
and exhibited almost perfect scaling for flat-MPI.
We studied cavitation phenomena with MD
involving up to 1.45 billion particles and observed multibubble
nuclei and Ostwald-like ripening.
Bubbles appear as a result of the many-body interaction of
particles, and Ostwald-like ripening appears as a result of the
many-body interaction of bubbles. This shows that
direct simulations of multiscale physics at the atomic scale are now possible.
In the case of more than 1 billion particles, the population dynamics of 
a bubble-size distribution can be observed with satisfactory statistics.
This will allow the study of the nonequilibrium dynamics of multibubble interactions.
We executed our codes on two different sites with different
methods of parallelization, thus showing that our codes are portable and flexible.
The flat-MPI method exhibited better performance than the hybrid,
while the computation assigned to each core was identical. Another issue is
to identify the main factor contributing to the parallel overhead of OpenMP.
The amount of memory per computational power on a node will decrease
as the number of cores per node increases. This trend will make it difficult to
perform simulations with flat-MPI in the future. The pseudo-flat MPI approach allows us
to perform a job with flat-MPI to obtain better performance
or with the hybrid to reduce memory consumption.
The developed code is available online~\cite{mdacp}.

\section*{Acknowledgements}
The authors would like to thank K. Nitadori, Y. Kanada, N. Kawashima, and S. Todo for valuable comments.
This work was partially supported by Grants-in-Aid for Scientific Research (Contract No.\ 23740287)
and by KAUST GRP (KUK-I1-005-04). 
The computational resource of Fujitsu FX10 was awarded by
``Large-scale HPC Challenge" Project, Information Technology Center of the University of Tokyo.
The computation was also carried out using the facilities of the Institute for Solid State Physics of the University of Tokyo, and Research Institute for Information Technology at Kyushu University.

\end{document}